\documentclass[10pt]{extarticle}
\usepackage[bordercolor=white,backgroundcolor=gray!30,linecolor=black,colorinlistoftodos, textsize=10pt]{todonotes}

\usepackage{amssymb}
\usepackage{textcomp}
\usepackage{amsmath}
\usepackage[textstyle,amssymb]{SIunits}
\usepackage{graphicx}
\usepackage[twocolumn,textwidth=183mm,columnsep=6mm,top=20mm,bottom=25mm]{geometry}
\usepackage{helvet}
\usepackage{titlesec}
\titleformat*{\section}{\bfseries\sffamily}
\titlespacing{\section}{0pt}{*4}{*0}
\titleformat{\subsection}[runin]{\normalfont\bfseries}{\thesubsection.}{3pt}{}
\usepackage{setspace}
\usepackage[breaklinks=true]{hyperref} 
\usepackage[font={footnotesize,sf},labelfont={sf,bf},labelsep=endash,justification=raggedright]{caption}

\usepackage{natbib}
\usepackage[labelfont=bf,justification=justified]{caption}
\usepackage{stfloats}
\usepackage{xr}
\begin{document}
\setlength{\emergencystretch}{3em}

\twocolumn[\begin{@twocolumnfalse}
	{\huge\sf \noindent\textbf{Femtosecond pulses from a mid-infrared quantum cascade laser}}
	\vspace{0.5cm}
	
		{\sf\large\noindent \textbf {Philipp Täschler$^{1*}$, Mathieu Bertrand$^1$, Barbara Schneider$^1$, Matthew Singleton$^1$, Pierre Jouy$^1$, Filippos Kapsalidis$^1$, Mattias Beck$^1$ and J\'er\^ome Faist$^{1,2}$}}
		\vspace{0.5cm}
		
		{\sf \noindent\textbf{$^1$Institute for Quantum Electronics, ETH Zurich, 8093 Zurich, Switzerland\newline}
				$^*$e-mail: {tphilipp@phys.ethz.ch}
				$^\dag$e-mail: {jerome.faist@phys.ethz.ch}}
		\vspace{0.5cm}
\end{@twocolumnfalse}]
\vspace{0.5cm}
\noindent{\sf \small \textbf{\boldmath
		 The quantum cascade laser (QCL) has evolved to be a compact, powerful source of coherent mid-infrared light. However, its fast gain dynamics strongly restricts the formation of ultrashort pulses. As such, the shortest pulses reported so far were limited to a few picoseconds with some hundreds of milliwatts of peak power, strongly narrowing their applicability for time-resolved and nonlinear experiments. Here, we demonstrate an approach capable of producing near-transform-limited sub-picosecond pulses with several watts of peak power. Starting from a frequency modulated phase-locked state, ultrashort high peak power pulses are generated via spectral filtering, gain modulation induced spectral broadening and external pulse compression. We assess their temporal nature by means of a novel asynchronous sampling method, coherent beat note interferometry and interferometric autocorrelation. These results open new pathways for nonlinear physics in the mid-infrared.
	}
}

The history of ultrashort lasers is tightly linked with the development of mode-locking\cite{Haus00}, whose ability of phase-locking longitudinal resonator modes using passive or active intracavity elements led to the demonstration of lasers with pulse lengths well below 10 fs\cite{Morg99, Sutt99}. For the first time, such sources allowed time-resolved measurements of ultrafast processes \cite{Hube01,Torr04}, revolutionised frequency metrology\cite{Udem02} and also find numerous commercial applications ranging from material processing to eye surgery. 

While ultrashort lasers have matured in the visible to near-infrared, large efforts are currently aimed at shifting ultrafast laser technology to longer wavelengths, to the so called molecular "fingerprint" region above 2.5 $\mu$m\cite{Pire15}. So far, ultrashort sources in this frequency range chiefly rely on down-conversion processes of shorter wavelength mode-locked lasers in nonlinear materials. Such sources suffer from low conversion efficiencies and often reach tabletop scale. Moreover, the damage threshold of the involved nonlinear crystals has become a limiting factor for generating high repetition rate high power laser pulses, as often desired for nonlinear spectroscopy applications\cite{Cao20}. The availability of compact sources which provide direct gain at these wavelengths, while also being compatible with mass production techniques, would greatly simplify existing laser systems.

Quantum cascade lasers\cite{Fais94} principally constitute ideal candidates for directly generating high-energy mid-infrared ultrashort pulses. They exhibit a low footprint, up to watt-level average power with a broad spectral bandwidth\cite{Jouy17, Schw17} and the peculiarity that their emission wavelength can be tailored by adapting the quantum well dimensions. Moreover, they have proved to be phase stable, a result of the strong third order susceptibility inside the gain medium, enabling the generation of phase-locked frequency combs\cite{Hugi12}. In contrast to typical mode-locked combs, QCL modal phases turn out to be mutually fixed but not constant, leading to a frequency modulated (FM) output instead of pulses\cite{Hugi12,Sing18}. This fundamental property of QCL combs can be ascribed to the short gain recovery time which is approximately one order of magnitude lower than typical cavity round-trip times\cite{Choi08}. 

Generating ultrashort pulses in such structures is an attempt of pushing the laser out of this favoured state. In the mid-infrared, active mode-locking was accomplished by strongly modulating a small section of the laser cavity close to the repetition frequency\cite{Wang09,Revi16, Hill20}. While available average powers were limited due to the onset of strong gain saturation, pulse durations as short as 3 ps were reported. Similar pulse lengths were demonstrated from actively mode-locked terahertz QCLs where mode-locking through gain modulation generally arises more naturally\cite{Barb11,Wang15,Wang17}.

In this paper, we follow a different path and take advantage of the unique properties of FM combs. Recent experimental\cite{Sing18,Hill19,Capp19} and theoretical\cite{Opac19,Burg20} works show that QCL combs emit a field of almost quadratic phase alongside a quasi-constant intensity. This finding is intriguing, as it means that well-established compression schemes can be applied for external pulse formation. Simultaneously, a close to maximally chirped intracavity field efficiently exploits the gain medium in place, principally allowing up to watt-level average powers\cite{Jouy17,Schw17}. Similar techniques are known from shorter wavelength semiconductor lasers using fiber dispersion for external phase compensation\cite{Chin93, Sato03, Rosa12}. 

\begin{figure}[t]
\centering
\includegraphics[width=88mm]{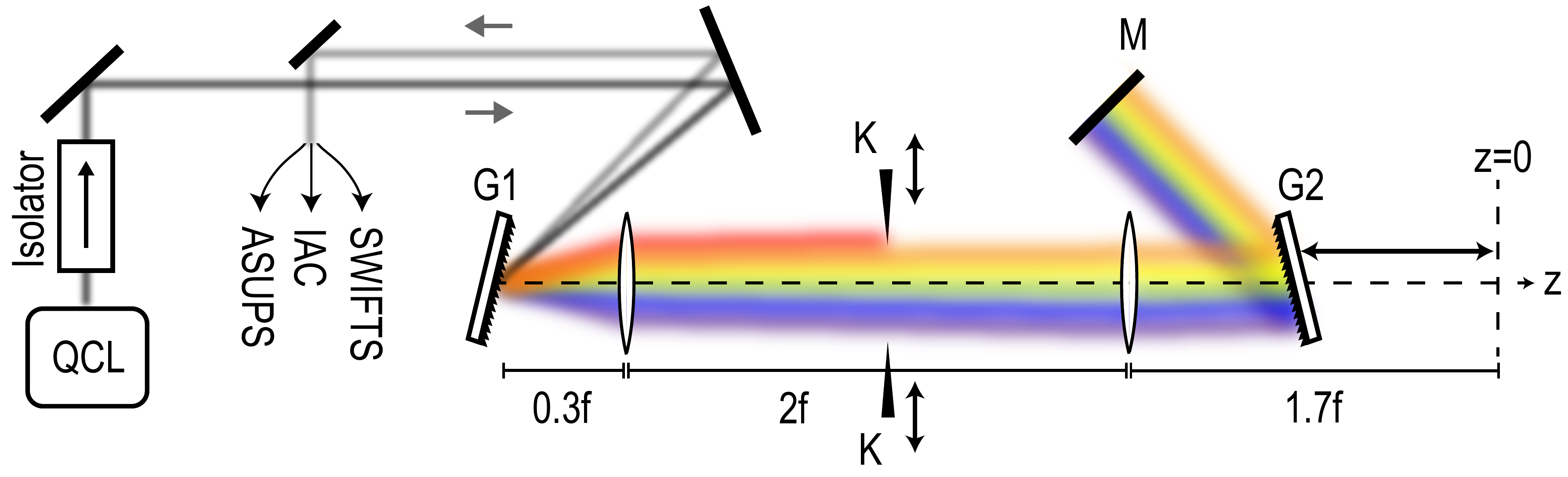}
\caption{\textbf{Diffraction grating compressor.} The collimated QCL beam is guided through a Faraday isolator before entering the compression step which is arranged in a reflective geometry. It is composed of two identical blazed gratings (G1, G2) with a 2f-imaging-system placed asymmetrically in between. The position of the second grating (G2) and the retroreflector (M) is adjustable allowing for precise GDD control. The optical spectrum can be spatially filtered in the centre focal plane of the compressor employing two knife edges (K). After compression, the pulses are characterised using asynchronous upconversion sampling (ASUPS), interferometric autocorrelation (IAC) and shifted wave interference Fourier transform spectroscopy (SWIFTS).}
\label{pulse_compressor}
\end{figure}

\section*{Results}
\noindent\textbf{Pulse compressor and coherent beat note interferometry.} The requirement of adding large amounts of positive group delay dispersion (GDD), of the order of some ps$^2$\cite{Sing18}, suggests the use of a grating compressor\cite{Mart87} as previously proposed \cite{Sing19}. The precise configuration utilised in this work is shown in Figure \ref{pulse_compressor}. By moving the second grating (G2) and retroreflector (M) with respect to the remaining setup the GDD can be gradually tuned. Even extended mode control is achieved by individually addressing spectral lines in the centre focal plane of the system using amplitude and phase masks.  Here, we restrict ourselves to a simple tunable spatial filter acting as a bandpass filter in Fourier space.

\begin{figure*}[b]
\centering
\includegraphics[width=180mm]{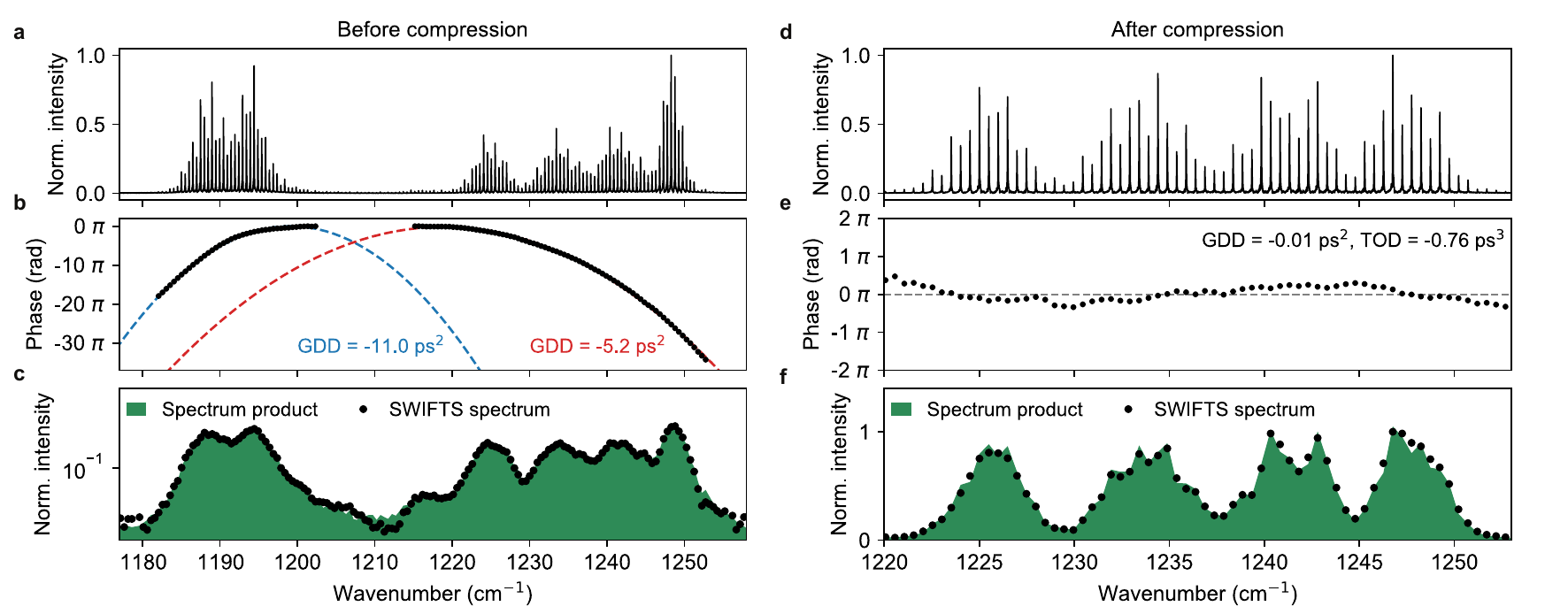}
\caption{\textbf{Complex comb spectrum and coherence before and after pulse compression as measured by SWIFTS.} \textbf{a} The comb spectrum emanating from the QCL consists of two distinct spectral lobes. \textbf{b} Within each lobe, spectral phases assume close to quadratic profiles which are well described by constant values of group delay dispersion (GDD). \textbf{c} The coherence of the frequency comb is exemplified by overlaying the SWIFTS spectrum with the spectrum product. \textbf{d} Using the tunable spatial filter incorporated in the pulse compressor, the lower wavelength spectral lobe can be isolated. The remaining amplitude spectrum covers a bandwidth of $\sim$25 cm$^{-1}$. \textbf{e} After pulse compression, the modal phases remain close to constant exhibiting a GDD of -0.01 ps$^2$ and a third order dispersion (TOD) of -0.76 ps$^3$. Considerable deviations from the flat phase profile are observed only at the edges of the intensity spectrum. \textbf{f} The intermodal coherence of the comb is maintained after pulse compression.}
\label{complex_comb_spectrum}
\end{figure*}

In this study, we use a 3 mm long QCL comb lasing around 8 $\mu$m featuring a plasmon-enhanced waveguide for dispersion compensation\cite{Jouy17}. Its repetition frequency is injection-locked\cite{Gell10} to an external stabilised radio-frequency (RF) source. In order to mitigate optical feedback, an isolator is placed at the output of the QCL before the beam enters the compression unit. A phase-sensitive measurement thereafter allows to monitor the complex QCL field. These measurements are carried out using coherent beat note interferometry, often referred to as shifted wave interference Fourier transform spectroscopy\cite{Burg14, Burg15} (SWIFTS).

The complex comb spectrum before compression is reported in Figure \ref{complex_comb_spectrum}a-c and shows a double-peaked amplitude spectrum. In accordance with previous experimental\cite{Hill19,Sing19} and theoretical\cite{Burg20} findings, these spectral lobes individually exhibit quadratic spectral phases with different GDD. The high degree of mutual coherence among comb modes, is, up to a constant prefactor, shown in Figure \ref{complex_comb_spectrum}c by overlaying the SWIFTS spectrum with the spectrum product (details in Methods).

From the emitted 580 mW of average power at the given operation point (details in Methods), a 100 mW compressed comb state is achieved. The compressor efficiency is evaluated to be $\sim$50\% and mainly limited by diffraction losses. Further losses are introduced by the Faraday isolator ($\sim$40\%) prior to the pulse compressor. Contrary to previous work\cite{Sing19}, phase compensation was only performed over the shorter wavelength spectral lobe, in order to maximally compress the given field. The tunable spatial filter was used for that purpose introducing spectral losses of $\sim$40\%.

\begin{figure*}[t]
\includegraphics[width=180mm]{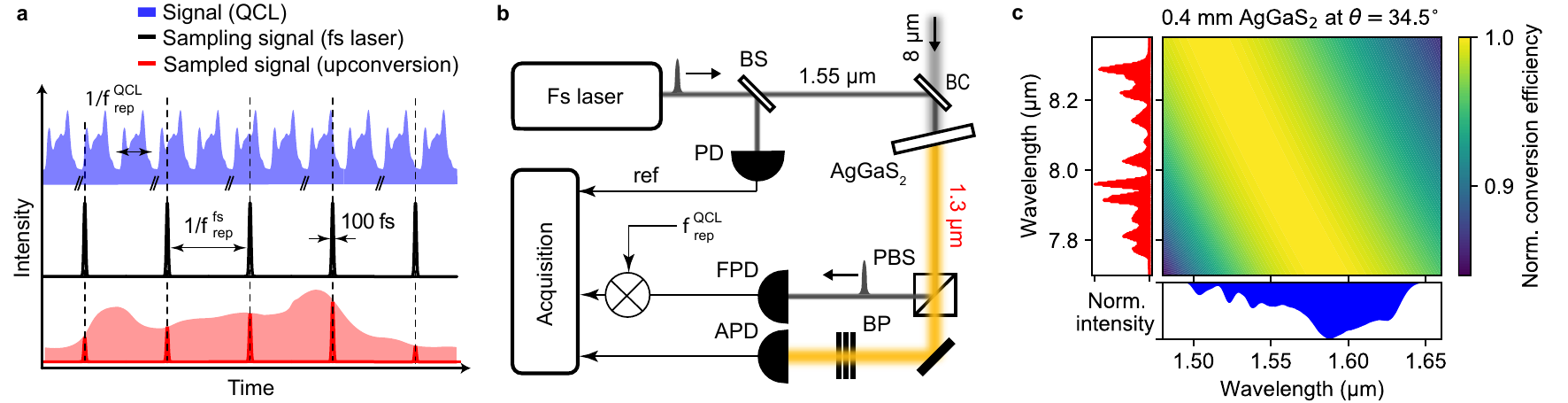}
\centering
\caption{\textbf{Asynchronous upconversion sampling.} \textbf{a} A femtosecond laser with repetition frequency $f_{rep}^{fs}$ is used to asynchronously sample the QCL waveform repeating at a round-trip frequency of $f_{rep}^{QCL}$. The envelope formed by the optical samples corresponds to one period of the QCL expanded along the time axis. \textbf{b} Experimentally, optical sampling is accomplished using sum frequency (SF) generation in AgGaS$_2$ and subsequent detection on an avalanche photodiode (APD), effectively intensity cross-correlating the fundamental fields. BS: beam splitter, BC: beam combiner, PD: photo diode, FPD: fast photo diode, PBS: polarising beam splitter, BP: optical bandpass filter. \textbf{c} The simulated normalised SF conversion efficiency as a function of the wavelengths of the contributing fields at the phase-matching angle $\theta$ is shown. Within the maximum spectral coverage of the QCL (red spectrum) and femtosecond laser (blue spectrum), we expect normalised conversion efficiencies $>$85\%.} 
\label{upconversion_sampling}
\end{figure*}

Figure \ref{complex_comb_spectrum}d-f shows the complex comb spectrum and coherence of the shorter wavelength spectral lobe after compression as measured by SWIFTS. Within a spectral bandwidth of 25 cm$^{-1}$, we observe a close to flat phase profile. Residual phase aberrations are limited to $\lesssim\pi$/4 and can no longer be accounted for by second order phase correction using the grating compressor. Considerable deviations in group delay are only observed at the edges of the intensity spectrum, where mode amplitudes are weak. 

In the following, we propose and experimentally demonstrate a novel cross-referencing method for characterising these compressed comb states. The nonlinear time domain technique allows for a full intensity reconstruction while setting stringent conditions on the phase coherence of the underlying waveform. We call this method asynchronous upconversion sampling (ASUPS).

\noindent\textbf{Asynchronous upconversion sampling.} The compressed intensity profile of the QCL is measured in an optical sampling experiment\cite{Taka93}, schematically described in Figure \ref{upconversion_sampling}a. The QCL waveform, repeating at a round-trip frequency f$_{rep}^{QCL}$, is asynchronously probed by 100 fs pulses from a mode-locked laser, whose repetition rate f$_{rep}^{fs}$ is approximately two orders of magnitude lower than f$_{rep}^{QCL}$. A slow detection system can thus be used to measure each of these samples and reconstruct the original QCL waveform once the ratio f$_{rep}^{QCL}$/f$_{rep}^{fs}$ is precisely known.

From an experimental viewpoint, optical sampling is achieved by frequency upconverting femtosecond pulses around 1.55 $\mu$m via sum-frequency (SF) generation in a nonlinear crystal \cite{Arge15, Kars05}, as shown in Figure \ref{upconversion_sampling}b. The resulting cross-correlation signal around 1.3 $\mu$m is proportional to the product of the intensities of the two incoming fields and can directly be measured on an avalanche photodiode (APD). The fundamental fields are spatially overlapped using a beam combiner (BC) and are collinearly focused on a 0.4 mm silver thiogallate (AgGaS$_2$) crystal meeting type-I phase-matching conditions. The crystal length is a compromise between achieving phase-matching over the full bandwidths of the involved fields, while maximising the generated SF signal. Figure \ref{upconversion_sampling}c shows the normalised conversion efficiency of the given crystal in dependence of the wavelengths of the fundamental fields, as calculated from its Sellmeier equations (details in Supplementary Section 1). The displayed QCL spectrum corresponds to its maximum spectral coverage attained close to current rollover.

The SF signal can be separated from the pump beams using a polarising beam splitter (PBS) as it is perpendicularly polarised. The remainder of the pumps and parasitic second harmonic generation in the crystal are further attenuated using a set of bandpass filters placed in front of the APD. The peak APD signal is then periodically sampled at the repetition rate of the femtosecond laser as optically measured on a photodiode (PD). In order to assign each sampling point to its QCL waveform position, the frequency difference $\Delta f = |f_{rep}^{QCL}-N\cdot f_{rep}^{fs}|$, where $N$ represents the index of the femtosecond laser beat note closest to $f_{rep}^{QCL}$, is simultaneously acquired. It is obtained by electrically mixing the high beat tone $N\cdot f_{rep}^{fs}$, as measured on a fast photodiode (FPD), with the repetition frequency of the QCL. 

\begin{figure}[tbh]
\centering
\includegraphics[width=88mm]{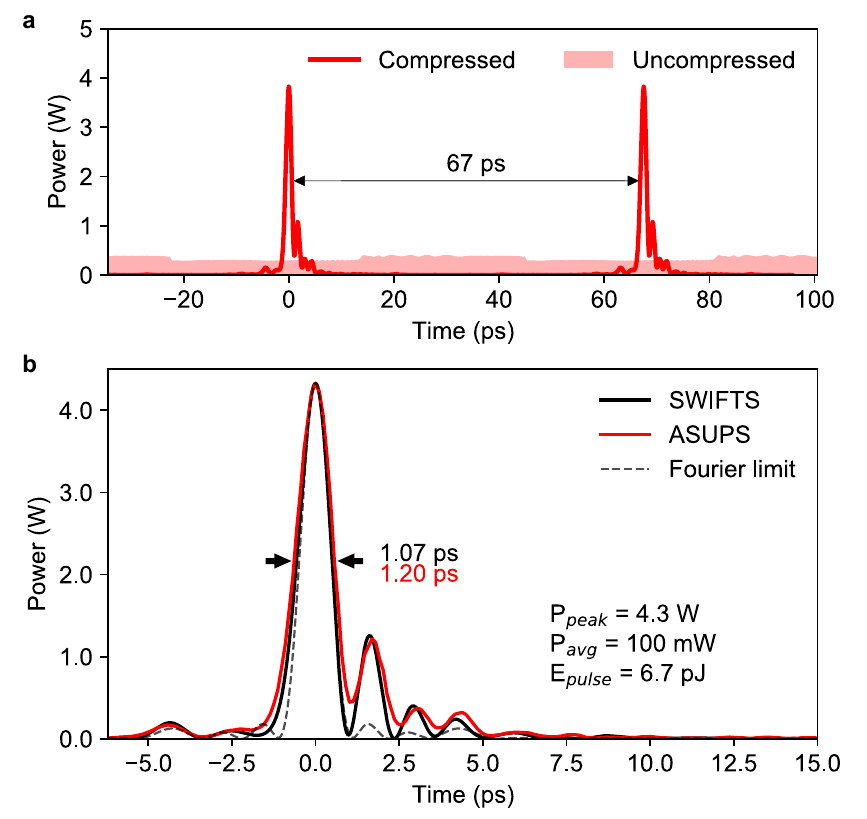}
\caption{\textbf{Compressed and uncompressed QCL intensity profile as measured by ASUPS.} \textbf{a} Before compression, we observe a quasi-constant intensity output. After compression, ultrashort pulses with a peak power of $\sim$4 W form. \textbf{b} Around the peak of the pulse, we compare the results obtained by ASUPS and SWIFTS, revealing a pulse duration of 1.20 ps and 1.07 ps, respectively. The transform-limit for the given spectrum lies at 0.98 ps.}
\label{pulses_upconversion_SWIFTS}
\end{figure}

The measured temporal intensity profiles for the states displayed in Figure \ref{complex_comb_spectrum} are shown in Figure \ref{pulses_upconversion_SWIFTS}a. Within one repetition period of 67 ps, the uncompressed QCL waveform remains quasi-constant, while a train of isolated pulses emerges after compression. When assuming the ASUPS pulse shape, the measured average power of 100 mW translates into a pulse peak power of $\sim$4 W. Further measurement data are enclosed in Supplementary Section 2. In Figure \ref{pulses_upconversion_SWIFTS}b, the measured ultrashort pulses are compared to the SWIFTS time domain reconstruction, normalised to the SWIFTS absolute power. From SWIFTS, we obtain a full width at half maximum (FWHM) pulse duration of 1.07 ps which is close to the Fourier limit for the given spectrum at 0.98 ps. A slightly increased pulse length of 1.20 ps is observed with ASUPS.

\begin{figure*}[hbt]
\includegraphics[width=180mm]{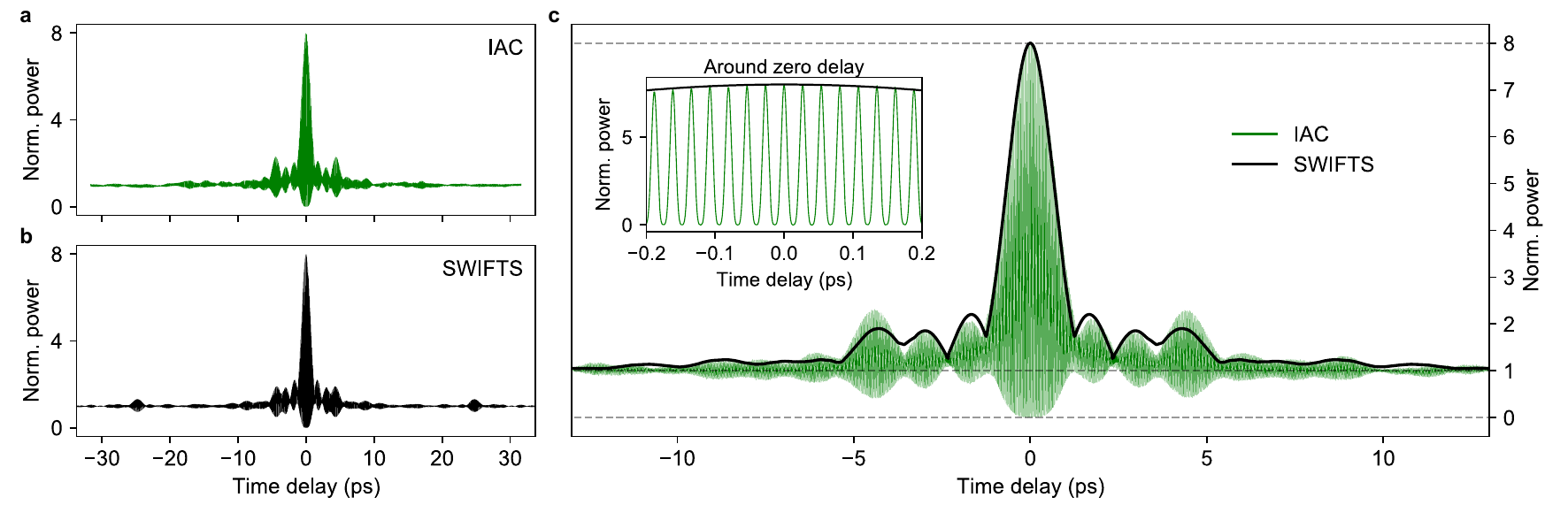}
\centering
\caption{\textbf{Compressed QCL pulses as measured by IAC.} \textbf{a} Two photon absorption in an InSb detector is used to directly record the IAC function of the compressed comb states. It exhibits a peak-to-background ratio of 8:1, featuring a sharp centre lobe with weak satellites in its wings. \textbf{b} The IAC function, as computed using the SWIFTS electric field, is shown. \textbf{c} The two interferograms are directly compared around the zero path difference, where, for better comparison, only the envelope of the SWIFTS IAC trace is plotted. The interferometric fringes around the zero delay are shown in the inset.}
\label{pulses_IAC_SWIFTS}

\centering
\includegraphics[width =180mm]{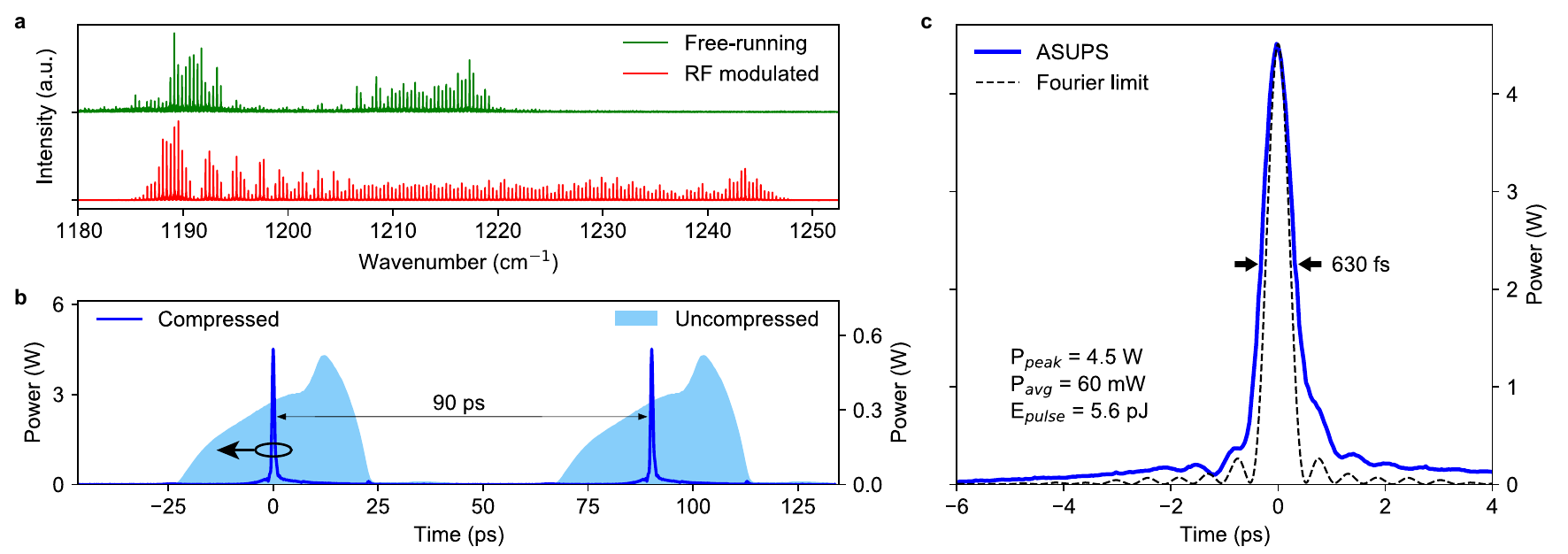}
\caption{\textbf{Shortest compressed QCL pulses.} \textbf{a} Optical spectra as obtained from the free-running and strongly RF modulated (35 dBm) QCL. Significant spectral broadening can be observed. \textbf{b} Measured intensity profile of the RF modulated QCL before and after pulse compression, where a train of ultrashort pulses is observed. The measurements were performed using ASUPS. \textbf{c} The obtained pulses exhibit a pulse length of 630 fs with a peak power of 4.5 W.}
\label{shortest_pulse}
\end{figure*}

Fully exploiting the high peak power and ultrashort pulse length of the femtosecond laser, ASUPS allows to reconstruct arbitrary periodic QCL waveforms, in particular also the important FM comb state\cite{Hugi12}. It is, however, ultimately limited in resolution by residual timing jitter between signal and sampling pulses\cite{Rodw86}. Autocorrelation measurements, on the other hand, are by definition phase coherent and are well suited for the characterisation of pulses. Yet, they only provide limited phase information and require considerable efforts to be applied to strongly chirped fields\cite{Picc19}.

\noindent\textbf{Interferometric autocorrelation.} As an additional means to confirm ultrashort pulse generation, we performed interferometric autocorrelation (IAC) using two photon absorption in an indium antimonide (InSb) photo detector\cite{Boik17}. Figure \ref{pulses_IAC_SWIFTS}a shows the obtained results. The IAC trace displays a sharp peak at the zero path difference which is approximately eight times higher than the background level. The wings of the IAC function feature small satellites and we observe interference patterns over the full length of the interferogram.

From the measured autocorrelation trace, the underlying pulse shape cannot be unambiguously extracted. However, from the SWIFTS electric field, we can explicitly calculate its corresponding IAC function and compare it to the measurement. The calculated IAC trace is shown in Figure \ref{pulses_IAC_SWIFTS}b and is compared to the direct measurement in Figure \ref{pulses_IAC_SWIFTS}c, where, for a more comprehensive view, we plot only the envelope of the SWIFTS IAC function. While observing small deviations at the interferogram's tails, excellent agreement of the two curves is observed at the centre lobe around the zero time delay.

\noindent\textbf{Femtosecond QCL pulses.} As indicated in Figure \ref{pulses_upconversion_SWIFTS}b, the ultrashort pulses reported so far are close to transform-limited. Consequently, in order to further decrease the pulse duration, the QCL optical bandwidth needs to be increased, while maintaining the quadratic phase profile. 

In this work, we employ strong gain modulation in specially designed RF QCLs\cite{Kaps21} for spectral broadening. A 4 mm long QCL with a microstrip-like line waveguide geometry is modulated at a power of 35 dBm close to its repetition frequency. Figure \ref{shortest_pulse}a shows the optical spectrum as obtained in free-running and RF modulated operation. In the latter case, we observe a significant increase in spectral bandwidth by approximately a factor of two. In time domain, such RF injection induces a strong overall amplitude modulation, as shown in Figure \ref{shortest_pulse}b. Due to increased gain saturation compared to free-running operation, the emitted average power is decreased in that case (details in Supplementary Section 3).

In the same subfigure, we report the temporal intensity as measured after pulse compression without filtering the QCL spectrum. We observe a train of ultrashort pulses with a repetition time of 90 ps. As can be seen in Figure \ref{shortest_pulse}c, these pulses are close to Fourier limited, exhibit a pulse length of 630 fs and a peak power of 4.5 W. Aforementioned measurements were carried out using ASUPS.

\section*{Discussion}\noindent
The measurements performed within the scope of this work unambiguously show that careful dispersion compensation of chirped QCL comb states allows the generation of ultrashort pulses. We proposed and experimentally demonstrated that further pulse shortening can be achieved via gain modulation induced spectral broadening.

Three mutually independent measurements, ASUPS, SWIFTS and IAC, reveal that these ultrashort pulses are temporally isolated. This finding follows directly from the reconstructed intensity profiles and the 8:1 peak-to-background ratio observed in the IAC trace. Moreover, the pulses are found to be near-transform-limited, as indicated by the pulse shape when compared to a pulse with constant phase profile and otherwise identical intensity spectrum. Further qualitative evidence comes from the observation of interferometric fringes over the full IAC interferogram, indicating only weakly chirped pulses. 

While exhibiting a similar pulse shape, ASUPS reveals a slightly increased pulse duration as compared to SWIFTS. We attribute this discrepancy to the combined effects of the finite femtosecond pulse length, dispersive walk-off inside the nonlinear crystal and uncompensated timing jitter among the fundamental laser fields, effectively limiting the method's time resolution to $\sim$100 fs. This is consistent with the IAC measurements which do not exhibit these limitations and are in accordance with the SWIFTS characterisation.

The absence of a constant intensity background in the ASUPS measurements is an experimental proof of the full coherence of the emitted waveform. This observation is underlined by SWIFTS, where the spectrum product and SWIFTS spectrum are commensurate over the full spectral bandwidth. It is to be noted that the proportionality constant relating the two quantities is not easily accessible by experiment. In IAC, the high degree of coherence manifests itself in the peak-to-background ratio which does not decrease even when looking at interference patterns of far away pulses, as reported in Supplementary Section 4.

Our work can be seen as a special case of chirped pulse amplification\cite{Stri85}, where it is the QCL itself which generates, naturally, a maximally chirped output with almost constant intensity. The emitted spectrum consists of segments of constant GDD, which can be isolated and externally compressed to form ultrashort pulses. From theoretical models\cite{Hill19,Burg20}, a close to constant GDD over the full laser bandwidth is expected for a more uniform gain spectrum. Such sources, in combination with recently demonstrated $\sim$100 cm$^{-1}$ spectral bandwidths\cite{Jouy17, Schw17}, could enable QCL pulses as short as 300 fs. Dispersion compensation could be achieved on-chip using low-loss integrated optics\cite{Bene17}. Moreover, strong gain modulation could serve as another means to further increase spectral bandwidths and correspondingly decrease pulse durations.

Similarly, with these spectral bandwidths and the reported $\sim$1 W average powers\cite{Jouy17, Schw17}, QCL sources with peak powers exceeding 100 W seem in reach. Improvements to reach these power levels may include using longer devices, with correspondingly lower repetition rates. Moreover, better feedback suppression could obviate the use of an optical isolator. These peak powers compare well with intensities needed to perform supercontinuum generation\cite{Cai15}, emphasising the potential of such sources also for broadband mid-infrared comb generation.

\section*{Acknowledgements}
\footnotesize
\setstretch{1.}
\noindent This work was supported by the BRIDGE program, funded by the Swiss National Science Foundation and Innosuisse, in the scope of the CombTrace (No. 176584; P.T., M.Ber., F.K.) project. Further financial support was provided by the Swiss National Science Foundation (No. 165639; M.S., P.J.) and the European Union's Horizon 2020 research and innovation program Qombs (No. 820419; B.S.). The author would like to gratefully thank J. Hillbrand for helpful advice and discussion while conducting the experiments and for proofreading the manuscript. Moreover, the author expresses his gratitude to S. Markmann and A. Forrer for their careful reading of the paper, S. Wang for his preliminary work on ASUPS and R. Wang for providing QCLs in an early stage of the work. We thank Dr. Emilio Gini of FIRST - Center for Micro- and Nanoscience for the MOVPE regrowths.
\normalsize

\section*{Author contributions}
\footnotesize
\noindent P.T. built the upconversion, SWIFTS and autocorrelation setup, carried out the experiments and wrote the manuscript with editorial input from M.Ber., B.S. and J.F.. M.Ber. characterised the normal buried heterostructure device (LIV, optical spectra) used for this publication, performed preliminary IAC experiments and helped with the setup of the RF optimised device. B.S. was involved in the SWIFTS analysis, characterised the RF optimised laser (LIV, optical spectra, beat note), helped with its setup and performed preliminary strong microwave modulation experiments. M.S. dimensioned the grating compressor. P.J. and F.K. processed the QCLs used in this work. M.Bec. was responsible for MBE growth. J.F. supervised this work. 
\normalsize

\section*{Competing interests}
\footnotesize
\setstretch{1.}\noindent
The authors declare no competing interests.
\normalsize

\section*{Methods}
\footnotesize
\noindent\textbf{Grating compressor:} The pulse compressor used in this study consists of two mirror-symmetrically placed Echelette gratings with a blaze angle of 35$^\circ$, 150 grooves/mm at an angle of incidence of 55$^\circ$. Two identical achromatic doublet lenses with focal lengths F=10cm are arranged in a 2f configuration and placed asymmetrically between the two gratings, a distance 0.3F away from the first grating (G1). The compressor is undispersive when the two gratings are separated by 4F. Synchronously moving the second grating (G2) and retroreflector (M) away from that position towards the first grating (G1), the introduced GDD linearly increases at a rate $\sim$0.5 ps$^2$/cm. Two knife edges, mounted on translation stages in the centre focal plane, enable spectral filtering. The incoming and outgoing beams are slightly detuned in propagation angle in order to spatially separate them with a pick-off mirror. For aligning the compressor, we used a beamprofiler at its output.
\newline
\newline
\noindent\textbf{Device:} The first QCL investigated in this paper is mounted epitaxial-side-down on an aluminium nitride submount which is kept at a constant temperature of -20$^\circ$ C using a Peltier element and a temperature controller. The device is high-reflection coated at the back facet and operated at a constant current of 0.60 A using a low noise current driver. The light-current-voltage (LIV) characteristic of the device and its optical spectrum as a function of bias current are shown in Supplementary Section 3. An external RF synthesizer emitting a power of 18 dBm is used to injection-lock the repetition frequency of the QCL via its bias line. These microwave powers do not induce a considerable change in the emitted waveform from the QCL as compared to the free-running case (details in Supplementary Section 5). A schematic of the injection circuit is reported in Supplementary Section 3. 

For the given device, pulse compression was additionally performed at a slightly higher bias current of 670 mA, where a spectral bandwidth of $\sim30$ cm$^{-1}$ is observed. In time domain, 850 fs pulses with a peak power of 7.4 W were measured using SWIFTS. Corresponding data can be found in Supplementary Section 3.

The second RF optimised device is mounted epitaxial-side-up and operated at a current of 0.95 A at a temperature of -25$^\circ$C. Its bias current is microwave modulated at 35 dBm via a coplanar probe placed at one end of the laser ridge. Further details are enclosed in Supplementary Section 3.
\newline
\newline
\noindent\textbf{Asynchronous upconversion sampling:} For the optical sampling experiment, we utilise a commercial erbium-doped mode-locked laser emitting $\sim$100 fs pulses at a repetition rate of 90 MHz and an average power of 240 mW. It is overlapped with the QCL beam using a longpass filter featuring a cut-off wavelength of 6.75 $\mu$m. The two beams are then collinearly focused on the AgGaS$_2$ crystal employing a parabolic mirror. After recollimation, the generated SF signal is separated from the pump beams with a polarising beam splitter and a spectral bandpass filter with a transmission window from 1.1 to 1.42 $\mu$m. 
\newline
\newline
\noindent\textbf{SWIFTS and IAC:} The SWIFTS and IAC setup is shown in Supplementary Section 3. It consists of a folded Mach-Zehnder interferometer with a QWIP, an MCT and an InSb detector at its output. 

For SWIFTS, the signal from the QWIP is quadrature demodulated in a lock-in amplifier using the electrically measured QCL repetition rate as a reference. From the in-phase and quadrature signals X($\tau$) and Y($\tau$), where $\tau$ stands for the time delay introduced by the interferometer, the analytic signal $Z(\tau)=X(\tau)-iY(\tau)$ can be computed. The phase differences of neighbouring comb lines $n$ and $n-1$ are then directly obtained from the argument of the Fourier transformed analytic signal $\phi_n-\phi_{n-1} = \arg{Z(\omega_n)}$. Cumulative summing allows to reconstruct the full modal phase profile $\phi_n = \sum_{m=1}^n(\phi_m-\phi_{m-1})$, starting from a randomly chosen phase $\phi_0$ which is of subordinate relevance as it only introduces a constant phase shift to the reconstructed electric field. Together with the spectral amplitudes acquired by conventional Fourier transform spectroscopy (FTS), the time domain field can be recovered. \\
This reconstruction scheme can only be applied for fully coherent frequency combs. In the context of SWIFTS, the coherence is typically assessed by comparing the SWIFTS spectrum which can formally be written as $|Z(\omega)|=|\left<E^*(\omega)E(\omega+\Delta\omega)\right>|$ to the spectrum product $\sqrt{\left<|E(\omega)|^2\right>\left<|E(\omega+\Delta\omega)|^2\right>}$ computed from FTS\cite{Methods_Burg14,Methods_Burg15}. Here, $E$ denotes the electric field, $\Delta\omega$ corresponds to the injection locked repetition frequency of the QCL and angle brackets represent averages over laboratory timescales.

For the IAC measurements, a photoconductive InSb detector is used instead which is operated as a two-photon detector. By recording its photocurrent as a function of the interferometer path delay, the IAC trace is obtained.

\normalsize
\section*{Data availability}
\footnotesize
\setstretch{1.}\noindent
The measurement data that support the plots within this paper are available from the corresponding author upon reasonable request. Data that support the findings in this Article are also available in the ETH Research Collection\cite{Methods_Tasc21}.
\normalsize

\section*{Code availability}
\footnotesize
\setstretch{1.}\noindent
The analysis codes will be made available upon reasonable request.

\end{document}